\documentclass{aa}  
\usepackage{graphicx}
\usepackage{balance}
\usepackage{txfonts}
 \newcommand\omicron{o}
\begin{document}

   \title{The extended molecular envelope of the asymptotic giant branch star $\pi^{1}$~Gruis as seen by ALMA}

   \subtitle{II. The spiral-outflow observed at high-angular resolution}

   \author{L. Doan
          \inst{1}
          \and
          S. Ramstedt\inst{1} \and W.~H.~T. Vlemmings\inst{2} \and S. Mohamed\inst{3,4,5} \and S. H\"ofner\inst{1} 
          \and 
          E. De Beck\inst{2} \and F. Kerschbaum\inst{6} \and M. Lindqvist\inst{2} \and M. Maercker\inst{2}  \and C. Paladini\inst{7} \and M. Wittkowski\inst{8}
          }

   \institute{Theoretical Astrophysics, Department of Physics and Astronomy, Uppsala University, Box 516, 751 20 Uppsala, Sweden\\
              \email{lam.doan@physics.uu.se} \and Department of Earth and Space Sciences, Chalmers University of Technology, SE-43992 Onsala, Sweden \and South African Astronomical Observatory, P.O. Box 9, 7935 Observatory, South Africa \and Astronomy Department, University of Cape Town, University of Cape Town, 7701, Rondebosch, South Africa \and  National Institute for Theoretical Physics, Private Bag X1, Matieland, 7602, South Africa \and Department of Astrophysics, University of Vienna, T\"urkenschanzstr. 17, 1180 Vienna, Austria  \and European Southern Observatory, Alonso de Cordova 3107, Santiago, Chile \and European Southern Observatory, Karl-Schwarzschild-Stra\ss e 2, 85748 Garching, Germany}

   \date{}

% \abstract{}{}{}{}{} 
% 5 {} token are mandatory

  \abstract
   {This study follows up the previous analysis of lower-angular resolution data in which the kinematics and structure of the circumstellar envelope (CSE) around the S-type asymptotic giant branch (AGB) star $\pi^{1}$~Gruis were investigated. The AGB star has a known companion (at a separation of $ \sim $400 AU) which cannot explain the strong deviations from spherical symmetry of the CSE. Recently, hydrodynamic simulations of mass transfer in closer binary systems have successfully reproduced the spiral-shaped CSEs found around a handful of sources. There is growing evidence for an even closer, undetected companion complicating the case of $\pi^{1}$~Gruis further.}
   {The improved spatial resolution allows for the investigation of the complex circumstellar morphology and the search for imprints on the CSE of the third component.} 
   {We have observed the $^{12}$CO $J$=3-2 line emission from $\pi^{1}$~Gruis using both the compact and extended array of Atacama Large Millimeter/submillimeter Array (ALMA). The interferometric data has furthermore been combined with data from the ALMA total power (TP) array. The imaged brightness distribution has been used to constrain a non-local, non-LTE 3D radiative transfer model of the CSE.}
   {The high-angular resolution ALMA data have revealed the first example of a source on the AGB  where both a faster bipolar outflow and a spiral pattern along the orbital plane can be seen in the gas envelope. The spiral can be traced in the low- to intermediate-velocity (13–25 km s$^{-1}$) equatorial torus. The largest spiral-arm separation is $\approx$5\farcs5 and consistent with a companion with an orbital period of $\approx$330 yrs and a separation of less than 70\,AU. The kinematics of the bipolar outflow is consistent with it being created during  a mass-loss eruption where the mass-loss rate from the system increased by at least a factor of 5 during 10-15 yrs.}
   {The spiral pattern is the result of an undetected companion. The bipolar outflow is the result of a rather recent mass-loss eruption event.}


 \maketitle
%
%________________________________________________________________

\section{Introduction}
The spiral structure of an extended AGB envelope was for the first time discovered in the carbon star AFGL 3068 \citep{maur06}. The unexpected morphology was explained when a companion was detected orbiting the primary \citep{morr06}. 
Similar circumstellar structures have since then been revealed around several AGB stars \citep[e.g.,][]{maer12,rams14,rams17,kim15}.  There have been a number of attempts to model the circumstellar morphology induced in a binary system with a mass-losing primary. \cite{moha12} proposed a mass transfer mechanism, referred to as 'wind Roche-lobe overflow' (wRLOF), in a detached binary system. If the dust condensation radius is comparable to the Roche-lobe radius, the mass transfer will form an accretion wake and
 the companion's orbital motion can shape the circumstellar envelope into a spiral morphology. \cite{chen17} studied mass transfer through Bondi-Hoyle accretion in binary systems with a low mass-loss-rate primary and reproduced the spiral structures found from observations. \cite{kim17} modelled recent detailed ALMA observations of the CSE around AFGL 3068 including an eccentric binary orbit. The outcomes of these simulations are affected by the accuracy of the equation of state, the opacity profile, and the grid resolution \citep{chen17}. 
 
 \begin{table}[th]
\centering
\caption{Summary of the ALMA observations used in this study. $  N_{ant}$ is the number of antennas}
\label{tab:observations}
\begin{tabular}{*{4}{c}}
\hline \hline
Array 				& On-source 			& $  N_{ant}$ 		& Baseline \\
        			   & time [min]                 		        			 &                   		& [m] \\
 \hline 
ALMA-ACA  				& 	23.7							 							& 			7				&  11-49 \\
ALMA 12-m  				& 	7.06													& 		31				&  15-438\\
              			&  7.06  												& 				41		& 15-348\\
ALMA-TP  					 & 	67.1													&  			3			& \\
 \hline
\end{tabular}
\end{table}

\begin{table}[th]
\centering
\caption{The spectral resolution $ \Delta \upsilon$ and the rms noise levels of the final images.}
\label{tab:images}
\begin{tabular}{*{3}{c}}
\hline \hline
LSR velocity				& $ \Delta \upsilon$  & rms \\
$ \left[ \rm{km\,s}^{-1} \right]  $		    &  $ \left[ \rm{km\,s}^{-1} \right]  $     & $ \left[ \rm{Jy\,beam}^{-1} \right]  $ \\
 \hline 
$ \left[ -40, 16 \right]  $	 &             2                   &  0.022 \\
$ <-40  $	 &              4                  &   0.006\\ 
$ >16  $	 &                  4              &  0.005 \\
 \hline
\end{tabular}
\end{table}

The role of a companion in shaping the AGB outflow has been investigated with our ALMA project where we have observed four well studied binary systems: R~Aquarii, Mira~($ \omicron$~Ceti), W~Aquilae, and $\pi^{1}$~Gruis. All  four systems show the imprint of companions on the AGB circumstellar envelopes. The smallest separation ($ \sim $10 AU) binary system of the Mira star R~Aquarii and its white-dwarf companion shows a ring-like structure aligned with the orbital plane \citep{rams18,buja18}. One possible explanation is that the structure is due to mass-loss variations. The CSE of the Mira~($ \omicron$~Ceti)  system, which has a moderate separation ($ \sim $60 AU), presents a complex morphology of bubbles and spiral arcs resulting from several dynamical processes during the evolution of the system. \citep{rams14}. The CSE of W~Aquilae and its main-sequence
companion at a separation of $ \sim $100 AU shows two weak spiral patterns with different inter-arm spacings \citep{rams17}. The larger-spacing pattern, which is brighter (denser) on the west side, can be reproduced from the known companion using an eccentric hydrodynamic wRLOF model of mass transfer. However, the smaller-spacing pattern is formed by an unknown process. The diverse morphologies seen in the binary systems reflect the complicated dynamic processes of AGB wind shaping. The outcome will depend on many parameters, e.g., the binary separation, the mass ratio, the wind speed of the mass-losing star, and the evolutionary stage of the companion.

\begin{figure*}
\centering
\includegraphics[width=\hsize]{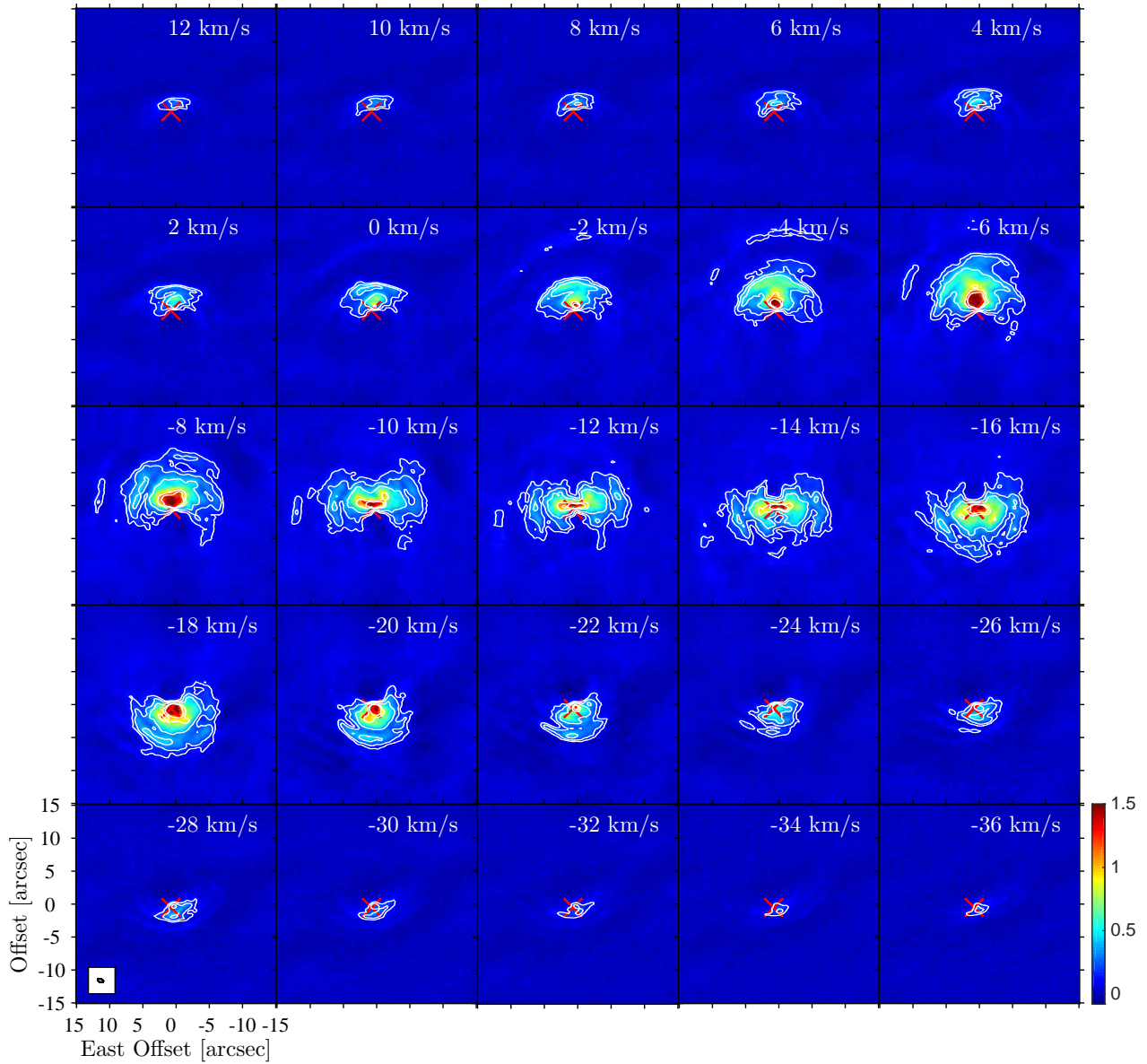}
\caption{Channel maps of the $^{12}$CO $J$=3-2 emission with the low-offset velocities. Contour levels are at 6, 10, 18, 40, and 50$\sigma$ ($\sigma$=0.022\,Jy\,beam$^{-1}$) and channel separation is 2\,km\,s$^{-1}$. The synthesized beam is plotted in the lower left corner of the -28\,km\,s$^{-1}$ channel. The local standard-of-rest velocity is given in the upper right corner of each channel. A cross denotes the stellar position determined from the continuum emission.  Colour code for the flux density (Jy\,beam$^{-1}$) is given in the colour bar.}
\label{fig:maps_lowV}
\end{figure*} 

\begin{figure*}
\centering
\includegraphics[width=\hsize]{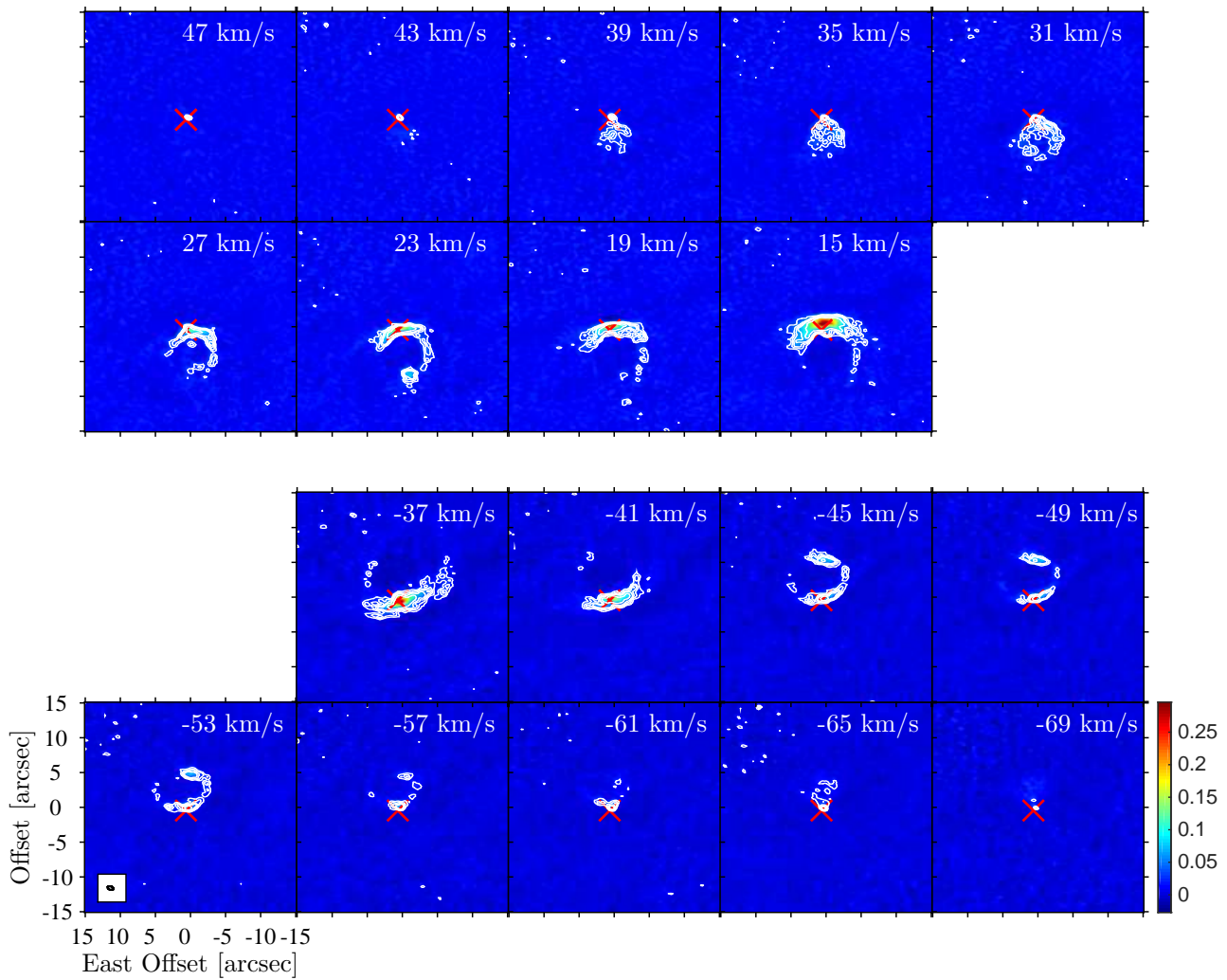}
\caption{Same as Fig. \ref{fig:maps_lowV}, for the $^{12}$CO $J$=3-2 emission with the high-offset velocities.  Contour levels are 3, 5, 7, 9, and 15$\sigma$ ($\sigma$=0.006\,Jy\,beam$^{-1}$) and channel separation is 4\,km\,s$^{-1}$. The synthesized beam is plotted in the lower left corner of the -53\,km\,s$^{-1}$ channel. The channels between 15 and -37\,km\,s$^{-1}$ are presented in Fig \ref{fig:maps_lowV}.} 
\label{fig:maps_highV}
\end{figure*} 

\begin{figure*}
\centering
\includegraphics[width=\hsize]{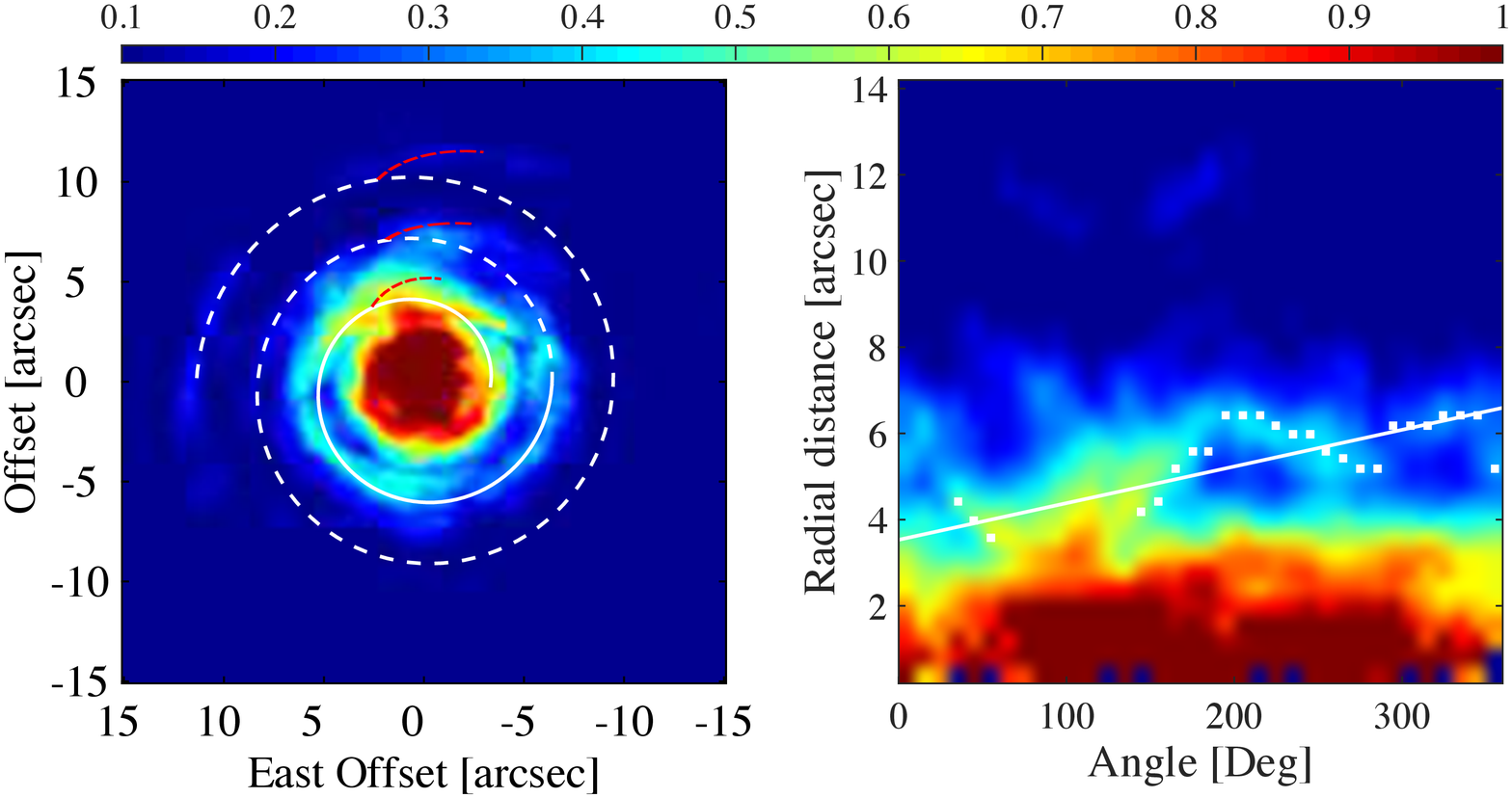}
\caption{ The image collapsed along the spectral axis (\textit{left}, see Sect. \ref{tab:images}), and the corresponding position-angle diagram (\textit{right}) with the peak positions marked by white squares, and a linear fit to the positions of the first winding. A spiral arm consistent with the fit is plotted on top of the image. A white, dashed part of the spiral is the continuation of the fit. Branches of the windings are marked by the red, dashed line. The flux scale is in Jy/beam.}
\label{fig:pos_angle}
\end{figure*}

The largest separation ($ \sim $400 AU) system in the project is the S-type AGB star $\pi^{1}$~Gruis and its main-sequence companion. The photosphere of the star appears round with no sign of Roche-Lobe shape \citep{pala18}.
The inner circumstellar envelope (within 5 stellar radii) probed by VLTI/MIDI appears very complex, and despite a first indication of spherical symmetry \citep{sacu08}, more recent observations \citep{pala17} tuned to measure asymmetries, show that the envelope is elongated. Observations probing larger scales revealed an asymmetric circumstellar envelope \citep{saha92,maye14}. A torus with a perpendicular bipolar outflow seen in the molecular gas was first mapped by \cite{chiu06} using the $^{12}$CO $J$=2-1 emission. The gravitational effect of the known companion on the AGB wind is not strong enough to focus
 material into a disk-like structure \citep{saha92}. Furthermore, the recent analysis of ALMA Atacama Compact Array (ACA) observations of the $^{12}$CO $J$=3-2 emission \citep{doan17} (Paper I) show that radiation pressure alone cannot provide sufficient momentum to drive the fast outflow. A closer
  companion has been suggested to explain the circumstellar morphology \citep{saha92}. Several possible shaping mechanism for the gas envelope were also discussed in Paper I.

This study presents the data analysis of the rotational line emission,  $^{12}$CO $J$=3-2 , from $\pi^{1}$~Gruis observed with the ALMA full array. The low-angular resolution data used in this study has already been discussed in Paper I. The data is used to constrain a 3D radiative transfer model to study the morphology, the temperature distribution, and the gas kinematics. Here the low- and high-angular resolution interferometric data, and the total power data are combined to investigate the circumstellar envelope with the hitherto highest angular resolution in this frequency range. The content of the paper will be presented in the following order: Observations and data reduction in Sect. 2; Observational results in Sect. 3; Discussion in Sect. 4; and Summary in Sect. 5.

\begin{figure*}
\centering
\includegraphics[width=\hsize]{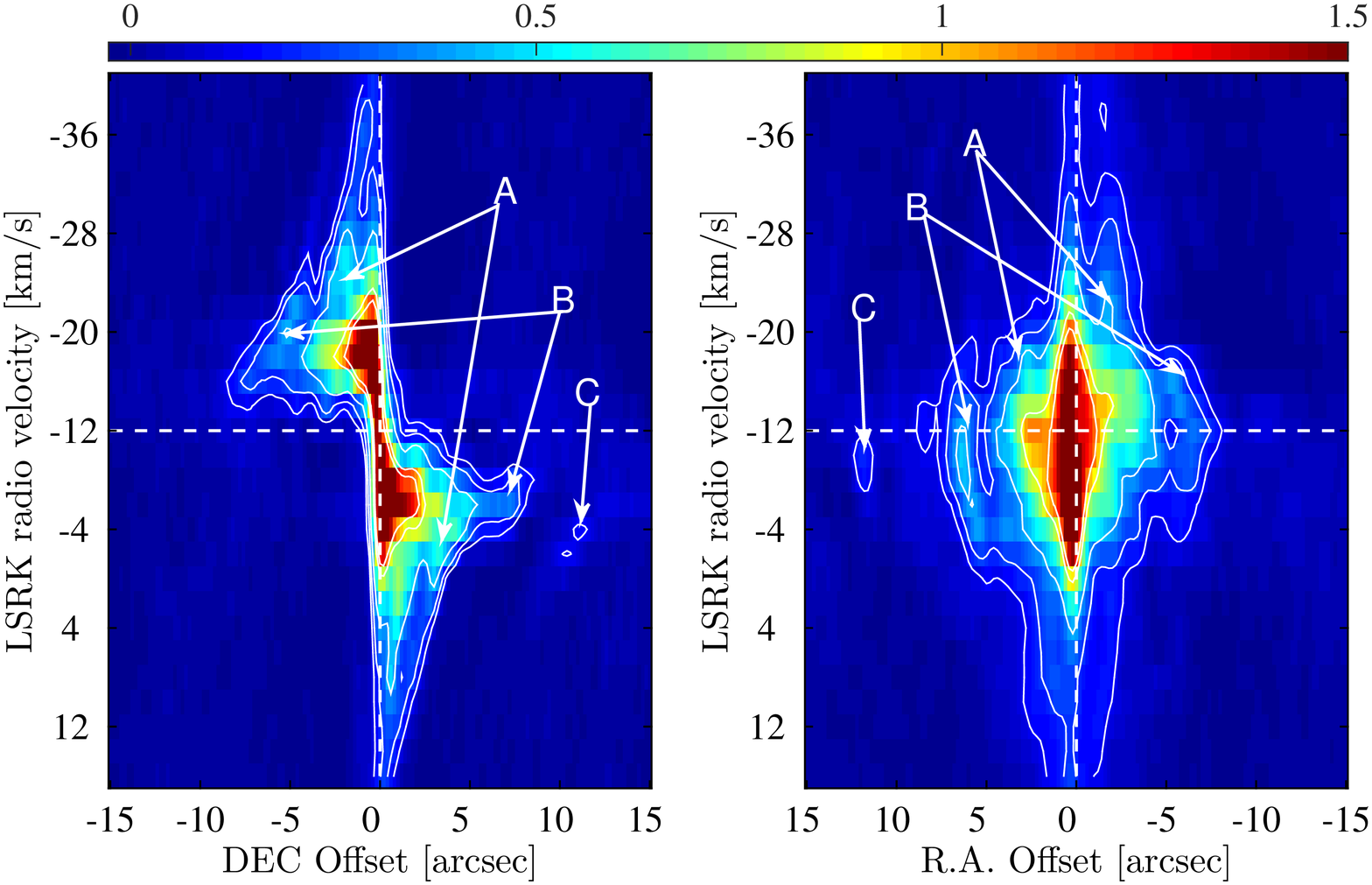}
\caption{Position-velocity diagrams along the DEC-axis (\textit{right}) RA-axis (\textit{left}) cut. The letters A, B and C indicate the first, second, and third observed winding, respectively. The color code and contour levels are the same as in Fig. \ref{fig:maps_lowV}. The vertical and horizontal dashed line show the stellar position and systemic velocity, respectively. The flux scale is in Jy/beam.}
\label{fig:pv_90}
\end{figure*} 

\section{Observations and data reduction}
\label{sec:obs}
We have observed $ \pi^{1} $~Gruis using both the compact array, ACA, and the 12-m main array. ALMA total power (TP) observations were also performed to complement the interferometric observations. A summary of  the observations is given in Table \ref{tab:observations}. The observations were done with four spectral windows that have a width of 2 GHz each, centered on 331, 333, 343, and 345 GHz.

The data reduction was done using the Common Astronomy Software Application (CASA 4.3.0). The interferometric data was first calibrated and combined separately,  then combined with the TP data. All the data was checked and adjusted for pointing positions to verify that the observations have the same phase center. %Because the on-source times of the two arrays were not optimal proportions for combining,%
The ratio between the ACA (7-m) array and the main (12-m) array weights was scaled to be 0.18. The combined data recovers  $ \sim$90\% of the flux observed by APEX in 2005.

The combined visibilities were finally imaged and deconvolved using the \textit{CLEAN} package using an iterative procedure with a decreasing threshold parameter. The initial spectral resolution of 0.5\,km\,s$^{-1}$ has been binned to 2 and 4\,km\,s$^{-1}$ for the low- and high-velocity channels, respectively, to improve the signal-to-noise ratio. The synthesized beam size is $0\farcs84\times0\farcs56$ at a position angle (PA) of $-$81$^{\circ}$. The properties of the final image cubes are summarized in Table \ref{tab:images}. 

\section{Observational results}
\subsection{Images}
\label{sec:images}
The velocity channel maps of the $^{12}$CO $J$=3-2 emission are given in Fig. \ref{fig:maps_lowV} and \ref{fig:maps_highV}. The emission at low-offset velocities (relative to the systemic velocity, $v_{sys}=-$12\,km\,s$^{-1}$) is dominating and shown separately from the high-offset velocity channels: the low-offset velocities (Fig. \ref{fig:maps_lowV}) with a channel separation of 2\,km\,s$^{-1}$ and high-offset velocities ((Fig. \ref{fig:maps_highV}))  with a channel separation of 4\,km\,s$^{-1}$. 

\textit{Low-offset velocity} ($\mid v - v_{sys} \mid < 25 $\,km\,s $^{-1}$) 

A striking feature in the channel maps is the central arc-like structure (Fig. \ref{fig:maps_lowV}). The structure was interpreted as a flared torus in previous studies \citep{chiu06,doan17} using lower-angular resolution data. Each channel with a certain line-of-sight velocity exhibits fragments of the arcs. Around the systemic velocity, $v_{sys}=-$12\,km\,s$^{-1}$, the fragments can be seen on both sides of the stellar position detected from the continuum emission (Paper I). At the systemic velocity channel, there are three arcs on the eastern side and two arcs on the western side. While arcs are distributed either to the north (in the red-shifted velocity channels) or to the south (in the blue-shifted velocity channels) at higher off-set velocities. The most distant arc from the stellar position, at  $ \backsim $12\arcsec,  is clearly seen at the 4\,km\,s$^{-1}$ channel. The arc separation appears to decrease and the arcs become indistinguishable at channels with velocities beyond 8\,km\,s$^{-1}$  and  $-$30\,km\,s$^{-1}$. 

Whether the arcs are sections of a spiral arm or concentric gas shells is further investigated in Fig. \ref{fig:pos_angle}. The left panel of Fig. \ref{fig:pos_angle}  shows an image collapsed along the spectral axis with the velocity range from $-$4\,km\,s$^{-1}$ to $-$20\,km\,s$^{-1}$. The pixel intensity at a given point in the right ascension (RA) - declination (DEC) plane represents the maximum intensity at that position in the spectral range. Fig. \ref{fig:pos_angle} (right) shows a diagram of the radial distance, $ r $, from the stellar position versus angle in polar coordinates, $ \theta $. The angle is zero at the west and increases counter-clockwise. The peak intensity points of the region $\pm$2\arcsec~from a radius of 5\arcsec have been fitted with a line in the radial distance-position angle plane in Fig. \ref{fig:pos_angle} (right). In Fig. \ref{fig:pos_angle} (left) a full spiral based on this fit is overplotted on the collapsed image. The starting point of the spiral cannot be exactly fitted since the central region is unresolved. Although the emission is weak in the outer parts, it seems consistent with a single (possibly disrupted) spiral arm density enhancement in the equatorial plane. The image, however, is a projection of the structure in the plane of the sky. Care must be taken to draw any further conclusion, e.g., the arm-spacing, since the spiral is rather complicated in reality. For example the inclination angle to the line-of-sight, the spatial extent in the vertical direction of the spiral, and the width of the spiral arm are not well constrained (Sect. \ref{sec:3Dmodel-discussion}). Closer inspection shows some branching of the spiral arcs. This consistently occurs at approximately the same position angle (to the north).  At each break there is an arc oriented in the opposite direction (see Fig. \ref{fig:maps_lowV} and the red, dashed line in the left panel of Fig. \ref{fig:pos_angle}). Along the spiral the northern windings appear brighter (than the southern), indicating some pile-up of material. Also, the southern part of the outermost winding is not seen in the observations.

\textit{High-offset velocity} ($\mid v - v_{sys} \mid > 25 $\,km\,s $^{-1}$) 

The shape of the fast bipolar outflow is mapped in Fig. \ref{fig:maps_highV}.  The high velocity outflow can be traced as two round bubbles at red- and blue-shifted velocities, e.g., in the 31\,km\,s$^{-1}$ and -53\,km\,s$^{-1}$ channels. The bubbles are incomplete circles at the other channels. This can be interpreted as two broken gas bubbles and/or a clumpy gas distribution.  The aligned bubbles are oriented perpendicular to the low velocity component, i.e., the bubble to the south is at red-shifted velocities (the first row) and the other is to the north at blue-shifted velocities (the last row). The most spatially extended emission is located at a radial distance of about $ \backsim $7\arcsec~ from the stellar position and seen in the 23 and 49\,km\,s$^{-1}$ channels. The average thickness of the bubbles is about 1\arcsec. The bubbles finally appear smaller at higher velocities, and then appear to be spots, e.g., in the 47\,km\,s$^{-1}$ and -69\,km\,s$^{-1}$ channels. There is no emission detected beyond these velocities. Therefore, the highest estimated velocity of the fast outflow is about 60\,km\,s$^{-1}$ relative to the systemic velocity.

\subsection{Position-velocity diagrams}
\label{sec:pv-diagram}

The correlation between the velocity field and the position vector in the low-velocity (between $-$40\,km\,s$^{-1}$ and 16\,km\,s$^{-1}$) channels is shown in Fig. \ref{fig:pv_90}. The position-velocity (PV) diagram is generated along the DEC-axis or at PA = 0$^{\circ}$ (\textit{right}), and the RA-axis or at PA = 90$^{\circ}$ (\textit{left}) cut using a 1-pixel wide slit. The three outermost windings of the spiral are marked with A, B and C. The winding C is not clearly seen on the southern or western side of the stellar position (e.g., also see Fig. \ref{fig:pos_angle}). This can be an effect of the real density distribution or the excitation conditions. The PV diagrams show two distinctive features. First, the windings appear most separated at the systemic velocity and converge toward the stellar position at higher off-set velocities. Second, in Fig.~\ref{fig:pv_90} (\textit{right}) the emission is only to the north in red-shifted velocities, and only to the south in blue-shifted velocities. These features can be interpreted as tracing a spiral which is tilted to the line-of-sight and which has a limited spatial extent in the meridional plane.

\section{Three-dimensional radiative transfer model}
\label{sec:3Dmodel-discussion} 
\subsection{Three-dimensional model}
\label{sec:model}

The gas envelope has been studied using a 3D radiative transfer model. The primary goal of the model is to investigate the morphology and the velocity field; therefore a detailed fit of the line intensity has not been attempted. The spatial distribution of the emission has been fitted with the highest accuracy hitherto attempted still within the constraints of equatorial plane reflection symmetry. 

\subsubsection{Morphology}
\label{sec:morphology}
The model consists of two components: a low-velocity spiral and a high-velocity bipolar outflow perpendicular to the spiral (Fig. \ref{fig:structure}). The spiral was constructed using a function,  $ R\sim \theta^{n} $, relating radial distance, $ R $, and angle, $ \theta $, in polar coordinates on the equatorial plane. The arm-spacing can be varied with different values of \textit{n}. The spiral is vertically flared with the height increasing with radial distance. The opening angle, $ \alpha$, of the spiral in the meridional plane is constrained by the spatial extent from the observed PV diagram. The bipolar outflow consists of two thin-wall bubbles radially expanding. The inclination angle between the line of sight and the equatorial plane is 40$^{\circ}$ as was determined in Paper I.

\subsubsection{Velocity field}
\label{sec:velocity}
\begin{figure*}
\centering
\includegraphics[width=\hsize]{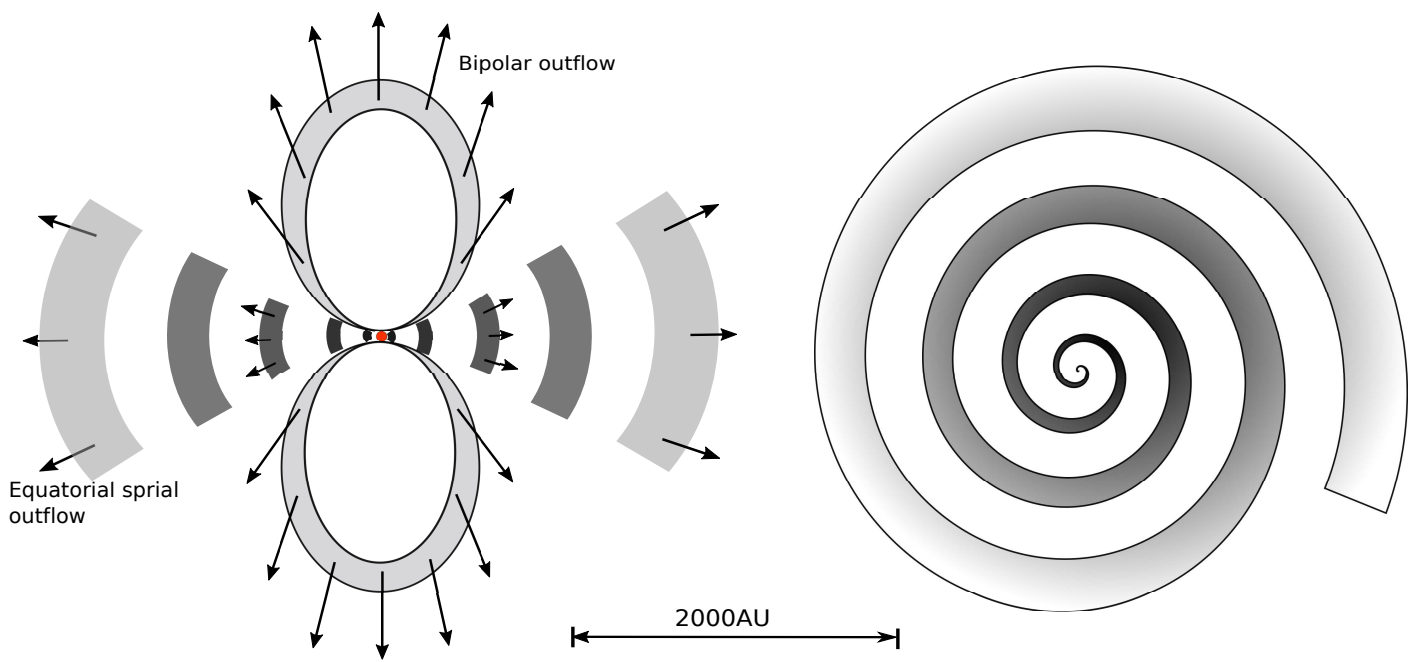}
\caption{Illustration of the CSE envelope viewed from the side (\textit{left}) and from the pole (\textit{right}). The central star, which is not included in the model, is marked by a red dot. The arrows represent the velocity vectors.}
\label{fig:structure}
\end{figure*}

%\textit{Velocity field}
The spiral is radially expanding as sketched in Fig. \ref{fig:structure}. The gas velocity law is adapted from the torus velocity from Paper I. The velocity depends on both the radial distance from the center, \textit{r}, and the latitude above or below the equator $ \varphi $, 
\begin{equation} \label{eq:velocity}
\centering
 v_{1}=\left[ 8+5 \frac{r}{R_{1}}\right]f_{\varphi},
\end{equation}
where $ R_{1} $ is the radial distance of the outermost edge of the spiral, the $ f_{\varphi} $ factor is unity at the equator and increases  linearly with latitude. It reaches a value of 1.9 at the top and bottom edges (which is smaller than the value of 2.5 used in the Paper I). The resulting maximum gas velocity is about 13\,km\,s$^{-1}$ in the equatorial plane and 25\,km\,s$^{-1}$ at the top and bottom edges. The bipolar outflow has a constant velocity of 60\,km\,s$^{-1}$. 

\subsubsection{Density and temperature distribution}
\label{sec:den_temp}
%\textit{Density and temperature distribution} 
The gas density functions for both the spiral and the inter-arm regions are adapted from the torus density in Paper I, 
\begin{equation} \label{eq:densty1}
\centering
n=A_{s,i}\left[ \frac{r}{10^{15} \rm{cm}}\right]^{-3} \frac{1}{f_{\varphi}},
\end{equation}
where $ A_{s} $ and $ A_{i} $ are respectively the scaling factors for the spiral and the inter-arm regions. The density contrast, $ \beta = A_{s}/A_{i}$, between the arm and the inter-arm regions within the spiral component can also be varied in the model. The $ f_{\varphi} $ factor is the same as in Eq. \ref{eq:velocity}. The outflow density is approximated by a simple inverse square law, 
\begin{equation} \label{eq:densty2}
\centering
n=B\left[ \frac{r}{10^{15} \rm{cm}}\right]^{-2},
\end{equation}
with the scaling factor $B$.  The gas temperature distribution is adapted from Paper I, $ T(r) = 190 (r/ 10^{15} \rm{cm})^{-0.85} $ and used for the whole structure. 

\subsubsection{Radiative transfer}
\label{sec:rt}
%\textit{Radiative transfer} 
The density, temperature, and velocity distributions of the envelope were constructed using the 3D modelling tool for complex structures, SHAPE \citep{stefetal11,santetal15}. The radiative transfer was performed using the non-LTE, non-local 3D radiative transfer code LIME \citep{brin10}. The excitation calculation is conducted for rotational energy levels up to $ J $=41 in the CO ground-vibrational state. The computational domain is a volume with radius of 2700~AU gridded in 100000 points. The calculation includes collisions with H$ _{2} $, and considers an ortho-to-para ratio of 3. The CO abundance relative to H$  _{2}$ is 6.5$ \times10^{-4} $ \citep{knap99}. The collisional rate coefficients, energy levels and radiative transition probabilities, are taken from \cite{scho05}. To take into account the effect of the interferometer and the atmospheric noise, the resulting images from the radiative transfer code were used to simulate the observations via the SIMALMA task in CASA using the same array configurations as for the observations. This task can simulate the combined data observed by the ACA (7-m), main (12-m), and total power arrays.  

\subsection{Modelling results}
\label{sec:model_rusult}
\begin{figure*}
\centering
\includegraphics[width=\hsize]{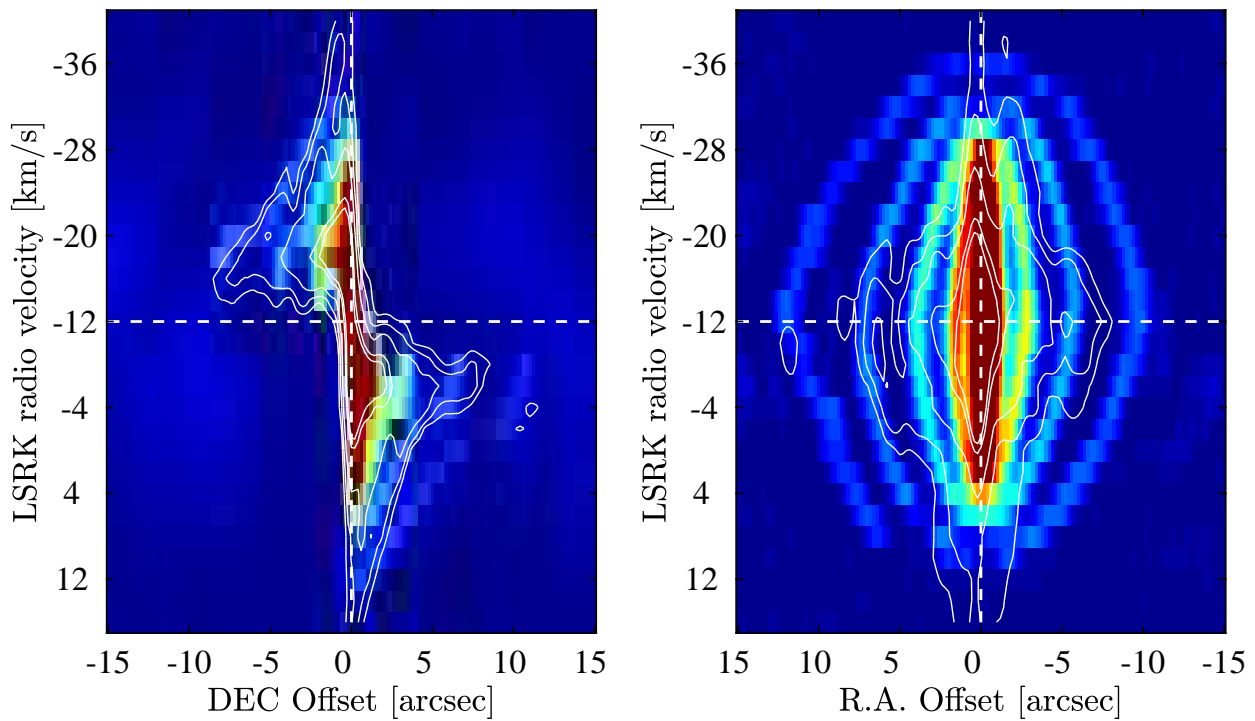}
\caption{Position-velocity diagram along the DEC-axis (\textit{right}) and RA-axis (\textit{left}) cut from the model. The contours from the observed PV diagram (Fig. \ref{fig:pv_90}) are overplotted.}
\label{fig:pv_model}
\end{figure*}

When modeling the full 3D structure, kinematics, temperature and excitation properties, traditional methods of finding a best fit to the data using e.g., model grids and e.g., $ \chi^{2} $-minimization quickly becomes insuperable. For this work, some of the best-fit parameters are selected from the analysis in Paper I (where a best-fit analysis to the line intensities and ratios was performed). The scaling factors in Eq. \ref{eq:densty1} and \ref{eq:densty2} for the density distribution of the spiral and the bipolar outflow are $ A_{s} = 1.8\times10^{8} $ cm$ ^{-3} $ and $ B = 1\times10^{7} $ cm$ ^{-3} $, respectively. For selecting the structure and kinematics described in Sect. \ref{sec:model} and this section, the model that could best reproduce the different features and structures seen in the channel maps and PV-diagrams was chosen by eye, but no formal fitting procedure has been performed. The modelling results were imaged in the same way as the observational data for analysis and comparison. The PV diagram in the model with $ \beta = 10$ is shown in the Fig. \ref{fig:pv_model}. It generally matches the observational PV diagram except for the outermost arc in the blue-shifted part in the left panel, or the rightmost arc in the right panel. The discrepancy is due to the missing part of the outermost winding of the spiral as mentioned in the Sect. \ref{sec:images}.  
 
The spiral arm is well-confined to the equatorial plane. The best fit value of the opening angle is about 40$^{\circ}$, which is larger than the value in Paper I, and dependent on the inclination. The angle cannot be larger than 50$^{\circ}$, otherwise the emission will be seen to, e.g., the north of the stellar position both in the blue- and red-shifted part (the left panel in Fig. \ref{fig:pv_90}). The spiral is less extended in the meridional plane than previously detected spirals, e.g.,  AFGL 3068 which has a fully extended spiral in the vertical direction \citep{kim17}. 

The spiral-arm spacing gradually increases outwards. The model with a spiral arm constructed from a $r\sim\theta^{2}$ function successfully reproduces the observed winding separation. The largest separation between two successive windings is $ \sim$5.5\arcsec \ measured at the outermost windings.          

The gas density contrast, $ \beta $, between the spiral arm and the inter-arm is not trivial to derive from modelling and an exact fit  has not been attempted. 
From the best fit model for the morphology and dynamics we have varied the quantity $ \beta $ to estimate its lower limit. 
Taking into account the effects of the interferometer, the lowest value of $ \beta $ in the model when one can discern the spiral from the gas envelope is approximately $ \beta_{min} \simeq 2 $. 
Appendix A  describes the method used to estimate the lower-limit  value of $ \beta $. 
The $\beta$-value required for detection will of course from case to case be dependent on the set-up of observations, e.g., the on-source time of observation, the noise level, and antenna configuration.    

\begin{figure*}
\centering
\includegraphics[width=\hsize]{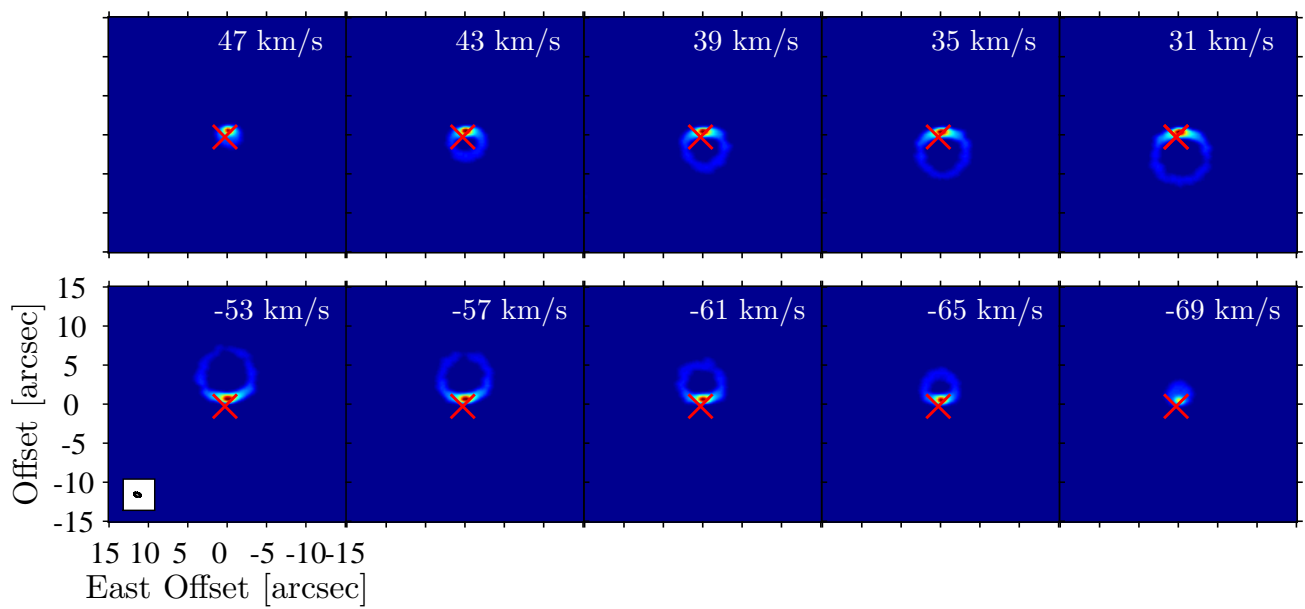}
\caption{The channel maps from the radiative transfer model showing the bipolar outflow at the high-offset velocities.}
\label{fig:maps_model}
\end{figure*}

The model can reproduce the high-velocity channels with a bipolar outflow that consists of two round, thin-wall bubbles with a constant radial expansion velocity of 60\,km\,s$^{-1}$ (see Fig. \ref{fig:structure}, left). 
The broken-bubble features seen in the observation are, however, not reproduced by model. As mentioned above, this is probably due to structure and/or clumpiness in the outflow and this level of detail has not been included in the model.
Fig \ref{fig:maps_model}  shows the high red- and blue-shifted velocity channels from the model. 
The emission region becomes smaller at higher velocities as seen from the observation (see Fig. \ref{fig:maps_highV}). 
The model also confirms that the observed spots at the highest velocity channels is not emitted from the caps of the outflow lobes as stated in Paper I, but from the gas that has the smallest angle between the velocity vectors and the line of sight.

\section{Discussion}
\label{sec:morphology}

\subsection{The morphology interpretation}
\label{sec:morph}

Hydrodynamical simulations of mass transfer in a detached binary system successfully reproduce a binary-induced spiral-shaped circumstellar envelope. According to \cite{mast98,mast99} the orbital motion of the mass losing star results in a spiral arm that can be more or less extended in the meridional plane. Another scenario emphasizes the role of the accretion disk created by material accreted  onto the companion. When the companion orbits the centre of mass, the disk sweeps through the dense envelope of the AGB star and shapes the envelope into a spiral structure.  The flattened spiral of $ \pi^{1} $~Gruis is likely created by a process similar to the latter case. With an arm spacing of $\Delta r\approx$5\farcs5 (see Sect. \ref{sec:images}) at a distance of 163 pc \citep{vanL07}, and an expansion velocity of $ v $=13\,km\,s$^{-1}$ (Paper I), the corresponding orbital period, $ T=\Delta r/v $, is 330 yrs. Assuming that the total mass of the system is 3\,M$_{\odot}$ \citep{sacu08}, the mean separation between the AGB star and the companion shaping the envelope should be about 70 AU. The period is about a factor of five larger than the period found from proper motion analysis \citep{maye14} with a companion at 10 AU separation. The results from the CO line data presented in this work supports the idea that a much closer companion, not the known companion with the period of 6200 yr and the separation of 400 AU, is responsible for the envelope structure. 

The brightness distribution of the CO lines might not exactly trace the envelope morphology. Around the systemic velocity, the $J$=3-2 emission shows a brightness peak close to the central position, while the $J$=2-1 emission has a two-peak (on either side of the stellar position) brightness distribution. \cite{chiu06} proposed that the gap between the two peaks is due to a central cavity associated with a dramatic decrease in the mass-loss rate over last 90 yr. In Paper I we showed that the difference seen between the CO $J$=2-1 and $J$=3-2 emission is due to a combination of optical depth effects and the exact morphology. From the analysis of the new higher-resolution  data we now conclude that the peak of the $J$=2-1 emission likely traces the higher gas density of winding A and that the central peak of the CO $J$=3-2 emission could correspond to a higher temperature region associated with the accretion disk around the companion.  

The analysis in Paper I showed that a constant expansion velocity of the torus component does not result in a good fit to the CO line profiles. A velocity that slowly increases outwards was chosen instead. This also agrees with the modelling results of the spiral structure presented in this paper. The lower expansion velocity results in the smaller arm separation in the inner region. However, the increase of the arm separation could also have an alternative explanation. The fast bipolar outflow is hypothesized to form in a short-duration eruption with a high mass-loss rate (see below). If the mass-loss eruption was isotropic and the gas collided with and accelerated the pre-ejected gas inside the spiral-shaped torus, this could result in an intermediate-velocity component. Since the mass-loss eruption must have occurred recently, the accelerated gas in winding A, would move outwards faster than the unaffected gas in the outer winding B. This would make the arm spacing smaller in the inner region of the envelope.

\begin{figure}
\centering
\includegraphics[width=\hsize]{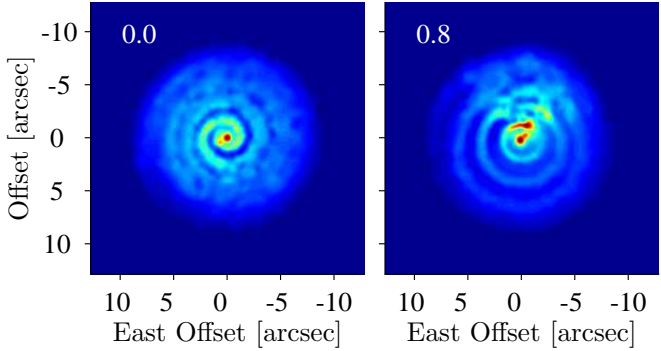}
\caption{The spiral structures from the hydrodynamic simulation for different orbital eccentricities. The eccentricity is given in the upper left conner of each panel.}
\label{fig:simulation}
\end{figure}

\subsection{The mass-loss eruption}
\label{sec:mass-loss}

The thin bubbles of the high-velocity component are likely created by a mass-loss eruption. The energetic and isotropic outflow is decelerated when colliding with the spiral-arm torus, while it is freely escaping in the polar regions and forms the bipolar outflow. From the thickness of about 1\arcsec, and the extent of about 8\arcsec, we estimate that the eruption occurred 100 years ago, and that the increased mass-loss rate lasted for approximately 13 years.
The gas mass released during the event can be estimated by assuming that the $J$=3-2 CO line emission is optically thin. 
 The optical depth for the CO $J$=3-2 line calculated in the radiative transfer model in Paper I is close to unity.
 The estimate includes both the mass in the bipolar outflow and the mass of the gas ejected into the spiral when the high mass-loss eruption occurred.  We applied the equation and procedure that
  was described by \cite{olof94}, and also used by \cite{chiu06} for the $J$=2-1 emission, 
\begin{equation} \label{eq:eqq3}
\centering
M= \frac{16\pi k m_{H}}{hc g_{u} A_{ul} f}\frac{D^{2}}{A_{e}} I Q(T_{ex})e^{E_{u}/kT_{ex}},
\end{equation}
where $ g_{u} $, $ E_{u} $ are respectively the statistical weight and energy of the upper level, $ A_{ul} $ is the Einstein coefficient, the CO abundance $ f = 6.5 \times 10^{-4} $, the distance  $  D = 160$ pc, the effective area $ A_{e}=\eta A_{g} $ with the aperture efficiency $ \eta = 0.76 $, and $ A_{g} $ is the
 geometrical area of the antenna. The integrated intensity $  I $ was first calculated over the high velocity range of the bipolar outflow, and then multiplied by a factor of 2. The factor of 2 accounts for the gas that moved into the spiral.  The partition function $  Q(T_{ex}) $ is approximately $(2kT_{ex})/(h\nu_{10}) $, where $  \nu_{10} $ is the $J$=1-0 line frequency. Owing to the high density (well above the critical density), the emission is satisfactorily thermalised. The excitation temperature $ T_{ex} = 15 $ K, which is also
  about the average kinetic temperature for a major part of the CSE, is adopted from \cite{buja97}. The estimated mass,  $ M $, is approximately $4.3 \times 10^{-5} $ M$_{\odot}$. The corresponding mass-loss rate is $3.3 \times 10^{-6} $~M$_{\odot}$yr$^{-1}$, which is five times larger than the average mass-loss rate before the mass-loss eruption (Paper I). Even though the radiative transfer model with the assumption of thin bubbles results is a good fit of the velocity of the walls, we cannot rule out a scenario were the walls are pushed outwards by a less dense wind. Moreover, since the calculated optical depth is about unity, the actual mass is probably underestimated and would be somewhat higher than the given value. Therefore, the estimated mass-loss rate should be considered as a lower limit.   

\subsection{Eccentric orbit of the undetected companion}
\label{sec:eccentric orbit}

The feature of the branching windings in the spiral arm (see Sect. \ref{sec:images}) was observed for the first time in the circumsteller envelope of an extreme C-star AFGL 3068 by \cite{kim17}. The authors suggested that a binary system with a high eccentricity of the orbit creates the feature. To investigate the formation of this particular feature, the observations are compared with a smoothed particle hydrodynamics (SPH) model of the circumstellar envelope in close binary system with an eccentric orbit. We have investigated effects of an eccentric orbit on morphology of the circumstellar envelop in a close binary system using a SPH simulation. The simulation uses the Evolved Stellar Interactions with GADGET in 3D that is a modified version of the GADGET2 code \citep[see, e.g.,][and references therein]{spri05,rams17}. The binary system consists of a 1 M$_{\odot}$  AGB star and a 0.6 M$_{\odot}$ companion in models with different eccentricities ($ e $=0.0, and 0.8). The binary separation is 10 AU and the AGB star has mass-loss rate of $2 \times 10^{-6} $ M$_{\odot}$yr$^{-1}$, and wind velocity of 11 km\,s$^{-1}$ (the terminal velocity of the wind in the absence of the companion). Fig. \ref{fig:simulation} shows the images of CO $J$=3-2 emission at the same LSR velocity from the models. The spiral is well defined in the circular orbit ($ e $=0.0) model. The branching feature seen in the observations is produced with an eccentric orbit ($ e $=0.8) model. The companion has a fastest speed and a smallest distance to the primary at the periastron passage. It strongly sweeps through the denser material region and drags a material stream along its motion direction. This results in the branching, clumpy pattern in the windings. The spiral is also brighter at one side (e.g., the periastron side) than at the opposite side. This can provide a good explanation for the feature seen in the $\pi^{1}$~Gruis envelope. This model ignores the torus-outflow structure of $\pi^{1}$~Gruis. A more detailed, self-consistent model of the spiral and bipolar outflow for $\pi^{1}$~Gruis will be studied in a future  paper. 

Understanding the formation of jets and/or bipolar outflows in AGB stars during the transition to planetary nebulae is still a challenge. The detected spiral structure is evidence for an undetected companion, however, it is not clear if the high mass-loss eruption can be triggered by this companion. \cite{saha16} describe the effect of a rotating accretion disk on a close companion in the carbon star V Hydrae and show that it is possible to launch a jet with velocity as high as the Keplerian velocity. However, the bubble-shaped outflow in $\pi^{1}$~Gruis is different from the bullet ejections observed in V Hydrae. Material flowing onto a companion can also heat the accretion disk and make a nova explosion, e.g., classical nova, or dwarf nova \citep{bode08}. The kinetic energy estimated from the measured velocity of the released mass is about two orders of magnitude smaller than total energy released in a typical classical nova. It is also possible that the high mass-loss eruption could be caused by the AGB star $\pi^{1}$~Gruis itself. 

\section{Summary}
\label{sec:sum}
High-angular resolution ALMA observations have revealed two important dynamical features in the CSE around $\pi^{1}$~Gruis: a circumstellar spiral pattern, as recently detected in several sources, observed together with a high-velocity bipolar outflow. Current hydrodynamical models study the formation of either of these two features. However, the observations of  $\pi^{1}$~Gruis presented in this paper now offers observational constraints to study their combined evolution and the impact that will have on the future shaping of a planetary nebula. In this paper, we present a model of the structure and kinematics of the CSE derived using the morpho-kinematical 3D code SHAPE in combination with the non-local, non-LTE 3D radiative transfer code LIME. We also discuss some of the specific features found in the spiral based on full 3D hydrodynamical modelling with post-processing radiative transfer (again using LIME). We find that the gravitation of the known companion is not strong enough to shape the envelope.  The orbital motion of a close-in, undetected companion (a third body in the system) is proposed to shape the dense dust-gas envelope into a spiral. The data analysis of the spiral arm has also shown that the companion is at a moderate separation of about 70\,AU where the accretion rate is probably not sufficient to drive the fast outflow. The properties of the bipolar outflow are consistent with being recently formed in a high mass-loss eruption. The mass-loss rate during the event increased massively (by at least a factor of 5) and later declined. 

\begin{acknowledgements}
The authors acknowledge support from the Nordic ALMA Regional Centre (ARC) node based at Onsala Space Observatory. The Nordic ARC node is funded through Swedish Research Council grant No 2017-00648. This paper makes use of the following ALMA data: ADS/JAO.ALMA\#2012.1.00524.S. ALMA is a partnership of ESO (representing its member states), NSF (USA) and NINS (Japan), together with NRC (Canada), NSC and ASIAA (Taiwan), and KASI (Republic of Korea), in cooperation with the Republic of Chile. The Joint ALMA Observatory is operated by ESO, AUI/NRAO, and NAOJ.\\
W.~H.~T. Vlemmings gratefully acknowledges support by ERC consolidator grant 614264.\\
S. Mohamed gratefully acknowledges the receipt of research funding from the National Research Foundation (NRF) of South Africa.\\
\end{acknowledgements}

\bibliographystyle{aa}

\bibliography{paper2}

%\begin{appendix}

\begin{appendix}

\section{The gas density contrast, $ \beta $, between the spiral arm and the inter-arm regions}
\label{app:find}
\begin{figure*}
\centering
\includegraphics[width=\hsize]{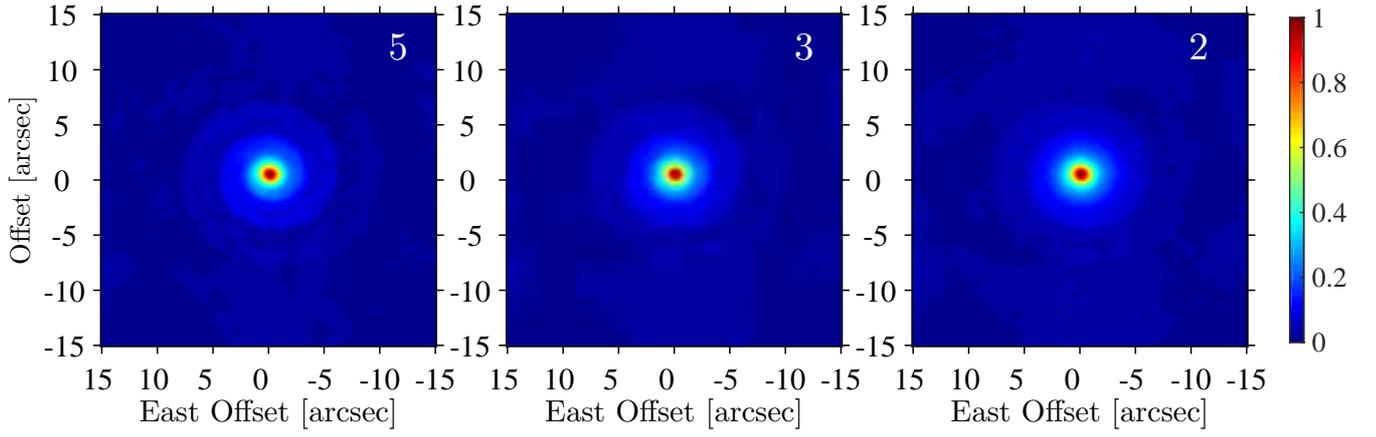}
\caption{The normalized-intensity images of the maximum intensity of each pixel across the spectrum with different values of $ \beta $, indicated in the upper right corner.}
\label{fig:den_cont_max}
\end{figure*}

\begin{figure*}
\centering
\includegraphics[width=\hsize]{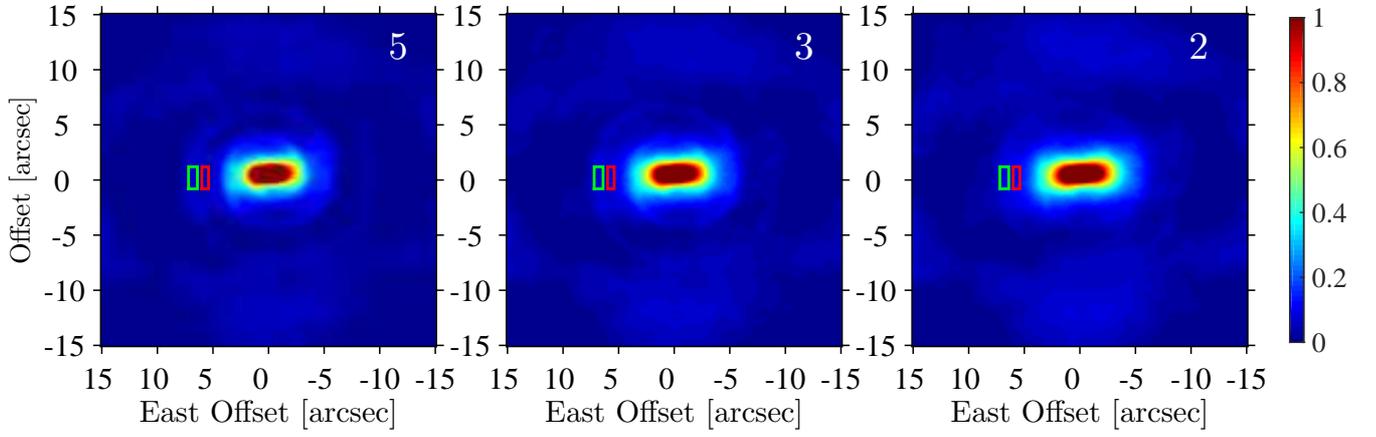}
\caption{The normalized-intensity images of the integrated intensity of each pixel across the spectrum with different values of $ \beta $, indicated in the upper right corner. The red and green rectangles, respectively, indicates the inter-arm region and the spiral arm as explained in the Appendix \ref{app:find}.}
\label{fig:integ}
\end{figure*}

\begin{figure}
\centering
\includegraphics[width=\hsize]{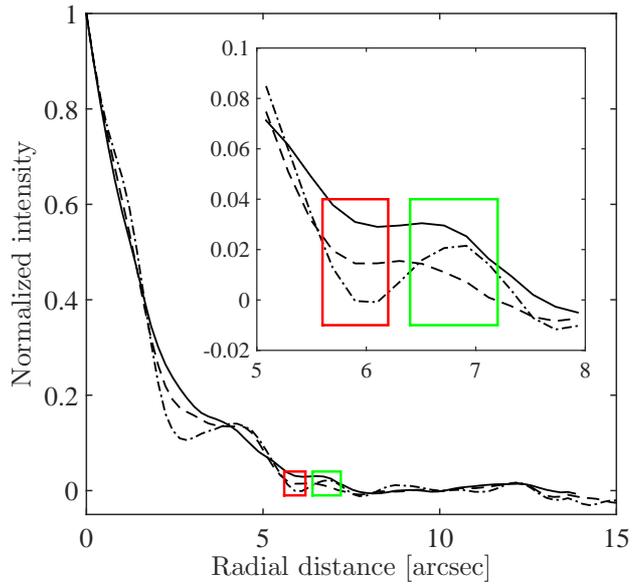}
\caption{The radial-intensity profile along the direction to the east in the models with the density contrast, $ \beta $ =5 (dashed-dotted line), 3 (dashed line), and 2 (solid line). The y-axis gives the normalized intensity averaged over 0.2\arcsec-wide rings within a $10^{\circ}$ circle sector. The red and green rectangles are the same as Fig. \ref{fig:integ}. The inset plot shows a zoom-in of the region around the rectangles.}
\label{fig:rad_pro}
\end{figure}

The density contrast, $ \beta $, is the ratio between the scaling factors $ A_{s} $ and $ A_{i} $, which are defined in Eq. \ref{eq:densty1}, for the spiral and inter-arm region, respectively. We assumed the same observational set-up as presented in the paper and estimated the lower limit of the density contrast for which the spiral pattern can be discerned with this set-up. In the radiative transfer model, $ A_{s} $ was kept constant and then $ A_{i} $ was varied to change the value of $ \beta $.  The output intensity distribution was then used to simulate the observation using CASA. The lowest value of $ \beta $ for which the spiral can be detected, is finally derived. We found that the resulting value is dependent on the $ A_{s} $ value. For example, in the model with a high value of $ A_{s} $, the gas excitation can be partly saturated and $ \beta $ must be really high to show a brightness contrast between the spiral arm and the inter-arm region in the image. Also, the resulting $ \beta $ value is dependent on the gas structure, e.g., the width of the spiral arms in the equatorial plane and the spatial extent of the inter-arm gas in the vertical direction. We used the same density function as the torus density in Paper I for the spiral and the gas structure in the best fit model for morphology and dynamics in Sect. \ref{sec:model}. Demonstrations of the models with different values of $ \beta $ are shown in the Fig. \ref{fig:den_cont_max} and \ref{fig:integ}. In the collapsed images showing the integrated intensity  in Fig. \ref{fig:integ}, we selected small areas within: the red rectangle indicating the inter-arm region and the green rectangle indicating the spiral arm. The rectangles are aligned along the direction to the east. If the root-mean-square intensity within the red rectangle is close to (>90\%) or higher than the maximum pixel intensity within the green rectangle, this means that the intensity gradually decreases with radial distance (Fig. \ref{fig:rad_pro}) and the spiral is barely discernible.  Fig. \ref{fig:rad_pro} shows the radial-intensity profile (produced using the CASAIRING\footnote{http://www.nordic-alma.se/support/software-tools} task in CASA) along the direction to the east in the model with the density contrast, $ \beta $ = 5, 3, and 2 (as in Fig. \ref{fig:integ}). The estimated lower limit of  $ \beta $ is approximately 2.

\end{appendix}

%\end{appendix}
\end{document}